\documentclass{moriond}
\usepackage{amsmath}
\usepackage{float}
\usepackage{hhline}
\usepackage{multirow}
\usepackage{pbox}
\def\Journal#1#2#3#4{{#1} {\bf #2}, #3 (#4)}

\def\PRL{\em Phys. Rev. Lett.}
\def\PRD{{\em Phys. Rev.} D}

\def\be{\begin{equation}}
\def\ee{\end{equation}}
\def\bea{\begin{eqnarray}}
\def\eea{\end{eqnarray}}

\newcommand{\Photo}{\includegraphics[height=35mm]{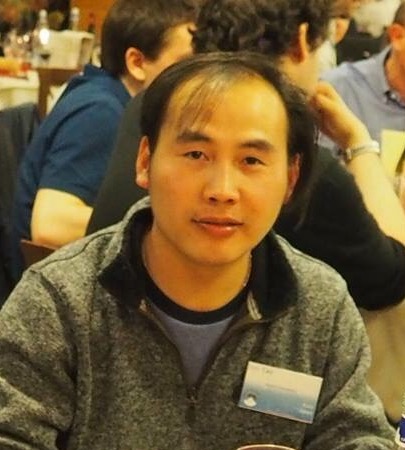}}
\newcommand\brabar{\raisebox{-4.0pt}{\scalebox{.2}{
\textbf{(}}}\raisebox{-4.0pt}{{\_}}\raisebox{-4.0pt}{\scalebox{.2}{\textbf{)
}}}}

\begin{document}
%\vspace*{1cm}
\title{Latest results from T2K}

\author{ S. V. Cao, on behalf of the T2K collaboration }

\address{Institute of Particle and Nuclear Studies, KEK\\
Ibaraki, Japan}

\maketitle
%and a total exposure of $2.65\times 10^{21}$ POT has been collected by the end of 2017
\abstracts{
Thank to the stable operation at intense beam power, T2K data with neutrino-mode operation almost doubled in one year. A number of critical improvements to the oscillation analysis have been introduced and resulted in an unprecedented level of sensitivity in searching for CP violation in the neutrino sector. T2K firstly reports that the CP-conserving values of parameter $\delta_{CP}$ in the PMNS mixing matrix fall out of its 2$\sigma$ C.L. measured range. 
}
%Along with this achievement, T2K has improved substantially the oscillation analyses by introducing a new selection, adding a new signal sample and understanding better the neutrino-nucleus interactions. All of these efforts result in more precise measurement of neutrino oscillation parameters and T2K firstly reports that the CP-conserving values of CP violation parameter $\delta_{CP}$ in the PMNS mixing matrix fall out of its 2$\sigma$ C.L measured range. To intensively explore CP violation in the lepton sector and continue to produce high-impact results, T2K proposes a program extension, T2K-II, to collect $20\times 10^{21}$ POT and ND280 upgrade is also planned to resolve uncertainties from neutrino-nucleus interaction modeling. Physics prospects of T2K-II and ND280 upgrade are sensational and T2K welcomes new collaborators for these developments.

%%%%%%%%%%%%%%%%%%%%%%%%%%%%%%%%%%
%%%%%%%%%%%%%%%%%%%%%%%%%%%%%%%%%%
\section{Introduction to Neutrino Oscillations }\label{sec:intro}
Neutrino oscillation phenomenon, in which neutrinos can transform from one flavour to another, indicates that neutrinos have mass and their flavor definitive eigenstates are different from their mass definitive eigenstates. This phenomenon is now well-established and so far is the only experimental evidence for the incompleteness of the Standard Model of elementary particles. Except for some known anomalies, the up-to-date neutrino data from both natural and artificial sources are well described by the $3 \times 3$ leptonic mixing matrix, so-called the PMNS (Pontecorvo, Maki, Nakagawa and Sakata) matrix \cite{Patrignani:2016xqp}. This unitary matrix, to connect the flavor definitive eigenstates, ($\nu_{e}, \nu_{\mu}, \nu_{\tau}$), with mass definitive eigenstates ($\nu_1,\nu_2,\nu_3$), is parameterized with three mixing angles ($\theta_{12}, \theta_{13}, \theta_{23}$) and one single Dirac phase $\delta_{CP}$\footnote{If neutrino is Majorana particle, two additional phases are included but should not take any impact on neutrino oscillation measurements.}, which represents the potential source of CP violation in the lepton sector,  as shown in Eq.~\ref{eq:osc33},
\begin{align}\label{eq:osc33}
\left(
\begin{matrix}
\nu_e\\
\nu_{\mu} \\
\nu_{\tau}
\end{matrix}
\right)
=
\left(
\begin{matrix}
1 & 0 & 0\\
0 &c_{23} & s_{23}\\
0 & -s_{23} & c_{23}
\end{matrix}
\right)
\left(
\begin{matrix}
c_{13} & 0 & s_{13}e^{-i\delta_{\text{CP}}}\\
0 &1 & 0\\
-s_{13}e^{i\delta_{\text{CP}}} & 0 & c_{13}
\end{matrix}
\right)
\left(
\begin{matrix}
c_{12} & s_{12} & 0\\
-s_{12} &c_{12} & 0\\
0 & 0 & 1
\end{matrix}
\right)
\left(
\begin{matrix}
\nu_1\\
\nu_2 \\
\nu_3
\end{matrix}
\right)
\end{align}
where $c_{ij} = \cos \theta_{ij}$ and $s_{ij} = \sin \theta_{ij}$.
Along with the above parameters, the probability for a $\alpha$-flavour neutrino to oscillate into $\beta$-flavour,   $P_{({\nu_{\alpha}}\rightarrow\nu_{\beta})}$, are driven by its energy, the distance it has travelled, and the mass-squared differences $\Delta m^2_{ij} = m^2_{i}-m^2_{j}$ where $i,j$ are mass eigenstate index. Remarkable efforts from various neutrino oscillation experiments for more than fifty years since its first experimental observation, the oscillation parameters, including three mixing angles and two mass-squared differences, are measured with high precision (within 5$\%$ uncertainty). However some crucial pieces are still missing to complete the neutrino oscillation picture described by the PMNS matrix, including the CP-violating phase $\delta_{CP}$ value, the mass hierarchy (sign of $\Delta m^2_{32}$) and whether or not $\theta_{23}$ is exactly $45^{\circ}$. Above all, the unitarity of the neutrino mixing matrix is still questionable.  
%$P_{\overset{\brabar}{\nu}_{\alpha}}$
%%%%%%%%%%%%%%%%%%%%%%%%%%%%%%%%%%
%%%%%%%%%%%%%%%%%%%%%%%%%%%%%%%%%%
\section{The T2K Experiment}\label{sec:t2k}
\begin{figure}[H]
\centering
\includegraphics[width=0.8\textwidth]{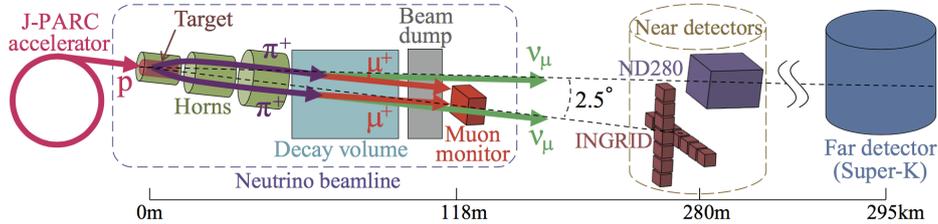} %histo
%histo
\caption{Schematic view of T2K experiment}
\label{fig:t2kschem}
\end{figure}

T2K (Tokai-to-Kamioka) is a long-baseline accelerator-based neutrino experiment placed in Japan. Details are described elsewhere \cite{Abe:2011ks}. % with neutrinos produced at  the Japan Proton Accelerator Research Center (J-PARC) and detected at Super-Kamiokande (Super-K) detector which play a role as T2K Far Detector. The 
Schematic view of T2K is shown in Fig.~\ref{fig:t2kschem}. The original goal of T2K is to discover the appearance of electron neutrinos from muon neutrinos, which had been completed in 2013 \cite{Abe:2013hdq}. T2K physic potential has been re-evaluated and the CP violation search in the lepton sector is put at the center of T2K targets since then \cite{Abe:2017vif}. In addition, T2K has a rich program of non-standard physics and neutrino interactions which are not covered in these proceedings. 

%--------------------figure
%--------------------figure
%\footnote{To achieve 1.3MW beam operation, it is required to be update to accept an intensity of $3.2\times 10^{14}$ protons per pulse with repetition cycle of 1.16 s. }
\noindent To collect data, T2K takes advantages of using one of the most intense and nearly pure $\nu_{\mu}$ ($\overline{\nu}_{\mu}$) beam from J-PARC. The 30 GeV proton beam with intensity of $2.5\times10^{14}$ protons per pulse (repetition cycle of 2.48 s) is extracted and bent $80.7^{\circ}$ toward the T2K far detector, Super-K, before striking onto a 90 cm-long graphite-cored target to produce charged pions and kaons. These mesons are focused or defocused by a system of three magnetic horns. The focused charge is defined by the horn current polarity, producing either a nearly pure $\nu_{\mu}$ or $\overline{\nu}_{\mu}$  from the focused secondaries decaying in the 96 m-long decay volume. Almost all hadrons except high energetic ($>$5 GeV) muons, which are monitored by so-called MUMON detector, are absorbed by a beam dump made of 3.2 m-thick graphite and iron places.  Several monitors are placed along the beamline to track the beam, monitor its (in)stability, measure number of protons passed through and to project its center position and direction right at the target. %All the information is used along with external data including the hadron production measurements from NA61/SHINE experiment to infer the neutrino fluxes for the T2K experiment. 
%The neutrino oscillation patterns are observed in the T2K far detector, Super-K, which is located 295 km away from the neutrino production target. 
%%Descibe ND280 here
There is a near detector complex placed 280 m from the neutrino production target, consisting of so-called INGRID and ND280 which are placed on-axis and 2.5$^\circ$ off-axis with the neutrino beam respectively. While the former is used to monitor the (in)stability of neutrino production and measure the neutrino profile around the beam center, the latter aims for characterizing the unoscillated neutrino beam and understanding neutrino-nucleon (nucleus) interactions. The central part of ND280 is the tracker which is composed of two fine-grained detectors (FGD1, 2) and three time projection chambers (TPCs). Both FGDs are used as the targets of neutrino interactions with  a total target mass of 1.1 tons each. While FGD1 is made solely of scintillator bars, FGD2 is scintillator-water interleaved. The most upstream of tracker places a $\pi_0$ detector (P0D) which consists of water bags placed between scintillator planes and brass or lead sheets. The P0D is optimized for tagging $\pi_0$, mainly coming from the neutral current process which can mimic the $\nu_e$ signal observed in T2K far detector.  Both tracker and P0D are surrounded by Electromagnetic calorimeters (ECals), which are designed to obtain high energy resolution and identification of neutral particles and electron/positron showers. Also ECals provide information whether particles are escaping or entering the tracker system. All sub-detectors are placed within a 0.2 T magnetic field produced by the former UA1/NOMAD dipole magnet. 
% \footnote{Water Cherenkov technique has limitation in separation neutrinos from antineutrinos}
%The ND280 has a sophisticated design with multi sub-detector providing excellent performance on classifying neutrino interactions based on the final state particles. 

\begin{figure}[H]
\centering
\includegraphics[width=0.4\textwidth]{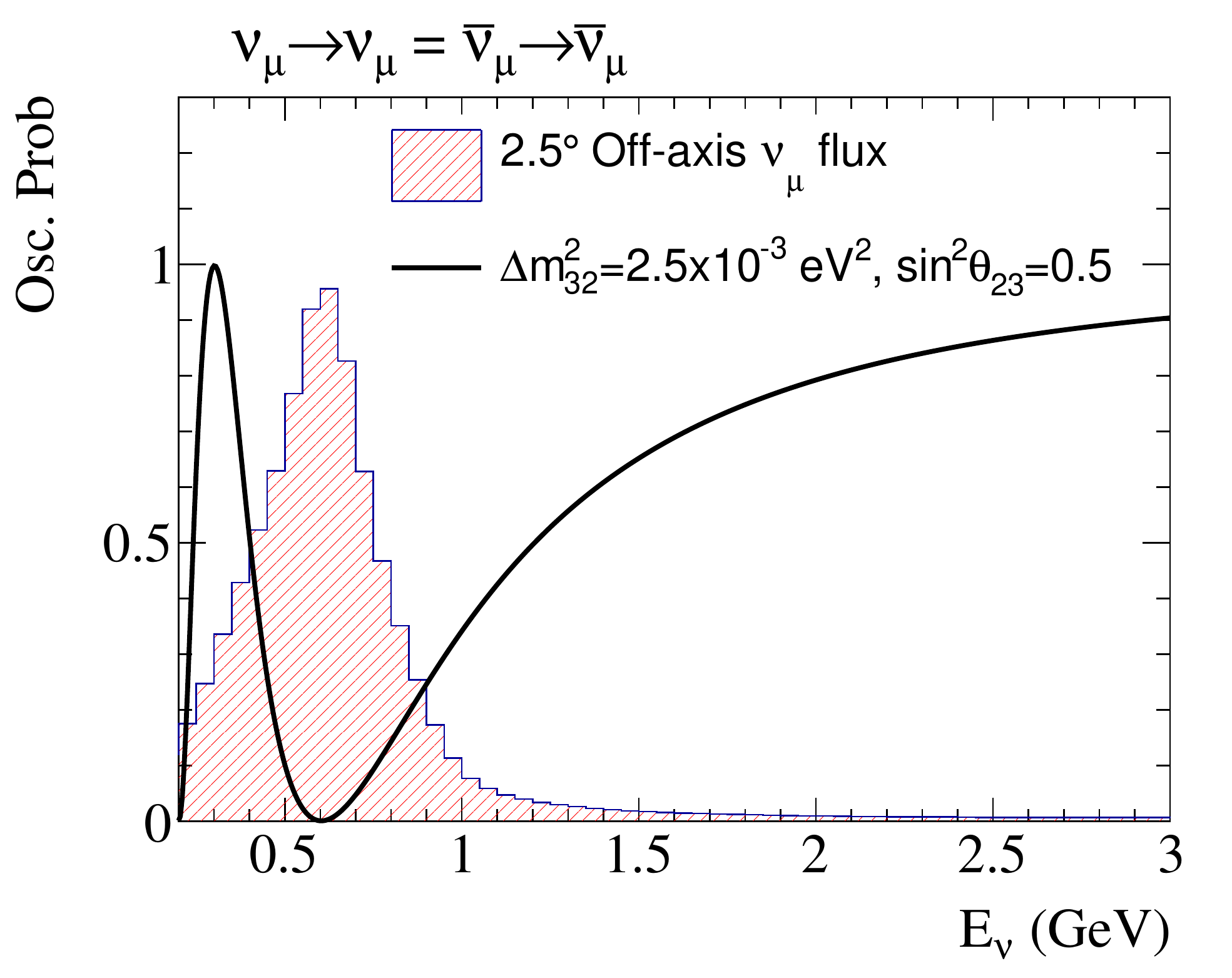} %histo
\includegraphics[width=0.4\textwidth]{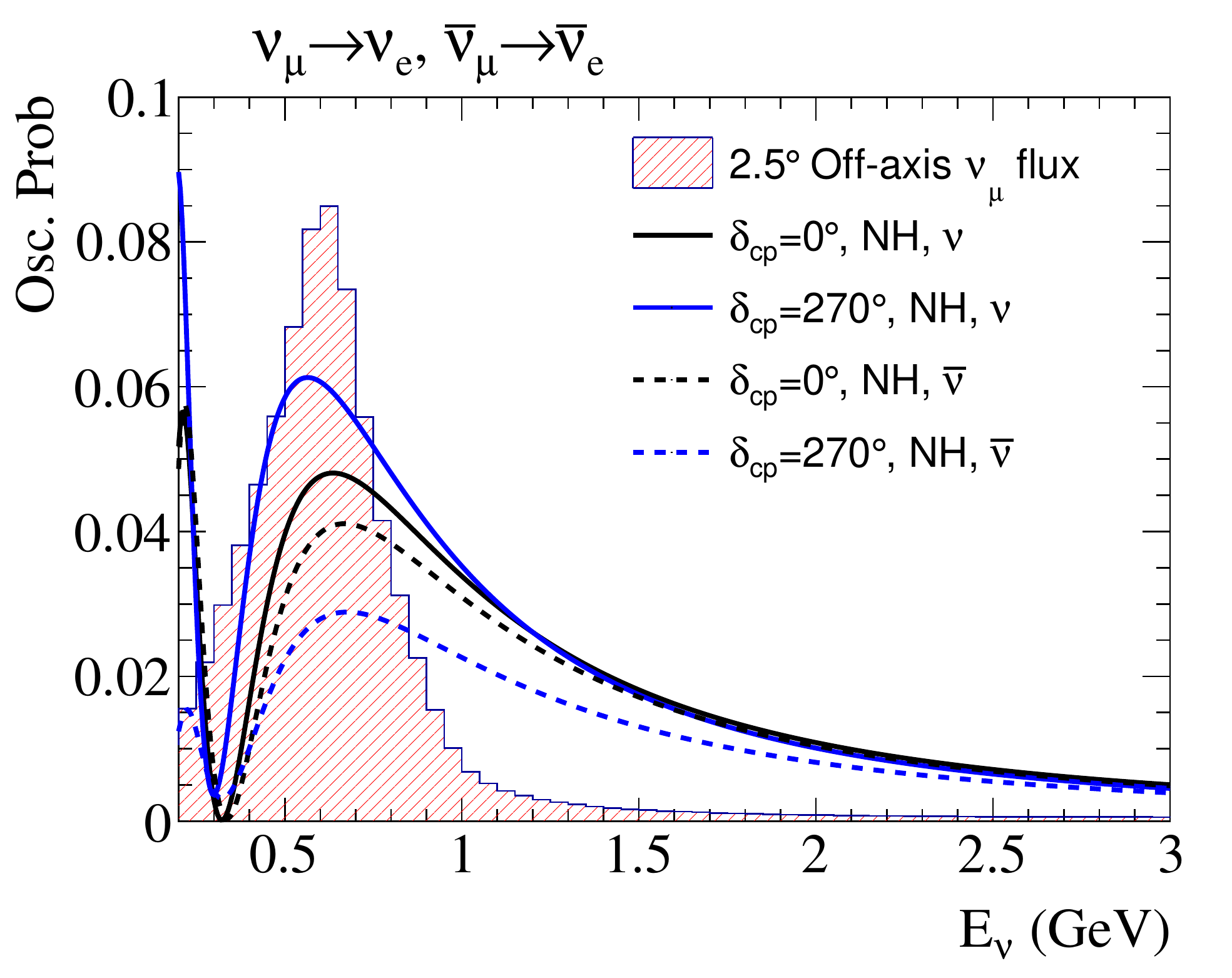} %histo
%histo
\caption{Oscillation probability of $\nu_{\mu}\rightarrow \nu_{\mu}$ (left) and $\nu_{e}\rightarrow \nu_{e}$ (right) with their antineutrino counterparts. }
%\caption{Oscillation probability of $\overset{\brabar}{\nu}_{\mu} \rightarrow \overset{\brabar}{\nu}_{\mu}$ on the left and $\overset{\brabar}{\nu}_{e} \rightarrow \overset{\brabar}{\nu}_{e}$ on the right}
\label{fig:t2kosc}
\end{figure}

 Excellent capability of Super-K detector in distinguishing muon and electron, consequently tagging $\overset{\brabar}{\nu}_{\mu}$ and $\overset{\brabar}{\nu}_{e}$, allows T2K to carry out two kinds of measurements: disappearance of $\overset{\brabar}{\nu}_{\mu}$ and appearance of $\overset{\brabar}{\nu}_{e}$ from the correspondingly genuine $\overset{\brabar}{\nu}_{\mu}$ beam. Fig.~\ref{fig:t2kosc} shows these oscillation probabilities with T2K flux overlaid. It shows that the narrow flux peak is right at the dip of the $\overset{\brabar}{\nu}_{\mu}\rightarrow \overset{\brabar}{\nu}_{\mu}$ and the peak of $\overset{\brabar}{\nu}_{e}\rightarrow \overset{\brabar}{\nu}_{e}$. This indeeds is resulted from the optimal design in which the Super-K is at an angular offset of 2.5$^\circ$ from the average beam direction.
%--------------------figure
%--------------------figure
%\footnote{Also to test CPT symmetry by comparing neutrino and anti-neutrino behaviours}
\noindent While the disappearance of $\overset{\brabar}{\nu}_{\mu}$ is essential to extract $\Delta m^2_{32}, \theta_{23}$ parameters,  the  appearance of $\overset{\brabar}{\nu}_{e}$ is sensitive to the $\theta_{13}$ and $\delta_{CP}$ parameters.  The CP violation if happen can manifest itself into the difference between the $\nu_{\alpha}\rightarrow\nu_{\beta}$ and its CP conjugate $\overline{\nu}_{\alpha}\rightarrow \overline{\nu}_{\beta}$.  In T2K, a complete three-flavor oscillation formulas including the matter effect with the earth density $\rho = 2.6 \text{ g/cm}^{3}$ is deployed. 
%\begin{align*}
%A_{CP} = P_{\nu_{\alpha}\rightarrow\nu_{\beta}}- P_{\overline{\nu}_{\beta}\rightarrow \overline{\nu}_{\alpha}}
%\end{align*} 
In practical, with conventional $\nu_{\mu} (\overline{\nu}_{\mu})$ beam, the CP violation phase can be measured by comparing $\nu_{\mu}\rightarrow\nu_e$ and $\overline{\nu}_{\mu}\rightarrow \overline{\nu}_e$ probabilities. In approximation at the maximum oscillation length of $\Delta m^2_{31}$, this CP asymmetry is expressed as: 
%equation 7.45, fumihiko suakane \cite{Suekane:2015yta} 
 \begin{align*}
 A_{CP}(\frac{\Delta m^2_{31}}{4E}L=\frac{\pi}{2}) = \frac{P_{\nu_{\mu}\rightarrow\nu_e}-P_{\overline{\nu}_{\mu}\rightarrow \overline{\nu}_e}}{P_{\nu_{\mu}\rightarrow\nu_e} + P_{\overline{\nu}_{\mu}\rightarrow \overline{\nu}_e}} \approx -0.27 \sin \delta_{CP} \pm \frac{L[km]}{2800}
 \end{align*}
where $+$  or $-$ sign depends whether mass hierarchy is normal or inverted respectively. For T2K experiment, with a relatively short baseline of 295 km, the mass hierarchy effect is relatively small, about 10$\%$; while the CP effect can be up to 27$\%$ when $\delta_{CP}$ is around $\pi/2$. %This is the most important factor for T2K to pursue the CP violation search in the leptonic sector which is shown in the next sections. 

%%Describe data set here
T2K started taking data from January 2010 and up to December 2017, a total $2.65\times 10^{21}$ Protons-on-target (POT), consisting of $1.51\times 10^{21}$ POT (57.14$\%$) in neutrino-mode operation and $1.14\times 10^{21}$ POT (42.86$\%$ in antineutrino-mode operation), have been delivered to T2K detectors. This exposure is equal to 34$\%$ of fully approved T2K data set. At the time of writing these proceedings, T2K is running in anti-neutrino mode with a stable beam power of 485 kW and a total exposure excesses $3\times 10^{21}$ POT.

%%describe data set here?
%%%%%%%%%%%%%%%%%%%%%%%%%%%%%%%%%%
%%%%%%%%%%%%%%%%%%%%%%%%%%%%%%%%%%
%\section{Detector Status and Physics Data Collection}\label{sec:det}
%Beam power steadily increased to 475 kW with high-quality data delivered. In total $2.65\times 10^{21}$ Protons-on-target (POT) delivered. However it is not all data covered for the result in this paper. In particular, $1.47\times10^{21}$ POT in neutrino-mode and $0.76\times10^{21}$ POT in anti-neutrino mode. \\
%The Super-K is 2.5$^\circ$ off the beam's average axis to achieve a narrow band beam peaked at oscillation maximum (0.6 GeV)
%With a muon and electron are well-separated thank to our well understanding of water Cherenkov light. \textcolor{red}{20 years operation of Super-K}
%%%%%%%%%%%%%%%%%%%%%%%%%%%%%%%%%%
%%%%%%%%%%%%%%%%%%%%%%%%%%%%%%%%%%
\section{ T2K Oscillation Analysis}\label{sec:ana}
As in previous T2K results \cite{Abe:2017vif}, neutrino oscillation parameters are estimated by comparing model predictions with observations at both T2K near and far detectors. The overall T2K oscillation analysis strategy is briefly described, following by crucial updates implemented for this reported result. The (anti-)neutrino fluxes are initially inferred by the combination of hadron production models, the external thin-target hadron-production data, the actual proton beam conditions, the horn currents, and the neutrino beam-axis direction measurements. At the peak energy ($\sim$ 600 MeV), the flux uncertainty is approximately 9$\%$, dominated by the uncertainty in hadron production data.  However, its impact on estimating the neutrino oscillation parameters, given that the T2K near detector and far detector are observing data from nearly the same flux, is largely suppressed when the T2K near detector data are included. In fact, due to the convolution of both flux and neutrino interaction models on the near detector data, the data samples from ND280 are used in the manner that the parameters modeling the flux and interaction models are constrained simultaneously. At last, the Super-K data along with its detector model are added in to the framework of oscillation parameter extraction. 
%, which directly affect on the number of signals observed at T2K far detector,
%\cite{Abgrall:2015hmv}
%Thank for stable operation at high beam power, the neutrino data set is doubled within one year. This statistic increase is the main driver for improving sensitivities on oscillation parameters. 
%Besides, this analysis uses new theoretical models to describe neutrino-nucleon (nucleus) interactions and a new event selection with the fiducial volume effectively extended for sampling signals in the Super-K data. 
%Particularly, T2K uses a event neutrino generator, so-called NEUT \cite{}, for simulating event-by-event interaction in the detector. The dominant channel at T2K energy range is the charged-current quasi-elastic (CCQE) interactions in which the neutrino-nucleon cross section pre-calculated from the Llewellyn-Smith model and the nuclear medium are described by the Smith-Moniz Relativistic Fermi Gas model.An improvement to the previous reports is that the effect of long-range correlations in nucleus is included.

Doubling the neutrino data set is the main driver for improving sensitivities on oscillation parameters reported in these proceedings. In addition to this, a number of critical improvements to the oscillation analysis have been introduced. Charged-current quasi-elastic (CCQE) interactions, dominating interactions at T2K energy range, which are simulated with primary neutrino-nucleon cross section calculated from the Llewellyn-Smith model and the nuclear medium described by the Smith-Moniz Relativistic Fermi Gas model, includes the effect of long-range correlations in nucleus.
 Furthermore, a model for multi-nucleon scattering which can mimic the CCQE-like signals\footnote{defined as events with lepton but no pion detected.} observed in T2K far detector when the knocked-out protons are below the water Cherenkov's threshold (about 1 GeV), are added in the simulation package. The contribution for this scattering to the CCQE-like sample is about 10-20$\%$ depending on the model. Processes of single incoherent pion production in the final state via the resonance excitation, originally described by the Rein-Sehgal model, are revived with the modified form factor and tuned to match with bubble chamber and most recent neutrino-nucleus data. For the single coherent pion production, a tuned model of Rein-Sehgal to match with the recent MINERvA data is deployed. %Following the inclusion of CC 1$\pi^+$ in T2K far detector samples, these processes are no longer just a background for oscillation analysis, but a signal.
%(from ANL, BNL experiments)  (from MiniBooNE and MINERvA experiments)

Another very important improvement in this analysis comes from the new reconstruction algorithm, called fiTQun, with complete charge and time information. This improvement allows to extend detector fiducial volume while keeping the selection performance at the same level to the previous reconstruction. This leads to 20$\%$ effective statistic increase in selecting e-like events. T2K also adds a brand-new signal sample for extracting the oscillation parameters, charged-current 1$\pi$ e-like sample which increases 10$\%$ statistics for neutrino e-like sample. In total, there are five far detector data samples used for extracting oscillation parameters, consisting of two samples of single-ring $\mu$-like events in both $\nu$-mode and $\overline{\nu}$-mode; two samples of single-ring $e$-like events in both $\nu$-mode and $\overline{\nu}$-mode; and the fifth one is single-ring $e$-like events with presence of one decay electron in $\nu$-mode. Except the fifth which mainly originates from the charged current single pion production, others are dominated by the CCQE interactions.
\begin{table}[H]
 \caption{Uncertainties on the number of predicted events in the Super-K samples from different systematic sources.}% Error on the ratio single-ring $e$-like events in $\nu$-mode to its in $\bar{\nu}$-mode, which affect directly on $\delta_{CP}$ measurement, is shown.}
\label{tab:syst}
 \begin{center}
   \begin{tabular}{l|c|c|c|c|c|c|}
     \hline \hline
                   \multirow{3}{*}{Uncertainty source}  & \multicolumn{6}{|c|}{$\delta_{N_{SK}}/N_{SK}$ (\%)}  
                   \\  \hhline{~------} 
                  &  \multicolumn{2}{|c}{ 1-Ring $\mu$} & \multicolumn{4}{|c|}{ 1-Ring $e$} \\ \hhline{~------} 
        & $\nu$ mode & $\bar{\nu}$ mode & $\nu$ mode & $\bar{\nu}$ mode & $\nu$ mode 1$\pi^+$ & $\nu$/$\bar{\nu}$  \\ \hline \hline
     SK Detector 		   &  1.86    & 1.51    & 3.03    &  4.22   & 16.69  &1.60   \\ \hline
     SK FSI $\&$ SI $\&$ PN  & 2.20   & 1.98  &  3.01  & 2.31  & 11.43 &1.57     \\ \hline
     ND280 constrained flux    & \multirow{2}{*}{3.22}  & \multirow{2}{*}{2.72} & \multirow{2}{*}{3.22} & \multirow{2}{*}{2.88} & \multirow{2}{*}{4.05}  & \multirow{2}{*}{2.50}   \\  \hhline{~~~~~~~} % updated
     $\&$ cross-section & {} & {} & {} & {} & {} & {} \\ \hline
     $\sigma_{\nu_{e}}/\sigma_{\nu_{\mu}}$, $\sigma_{\bar{\nu}_{e}}/\sigma_{\bar{\nu}_{\mu}}$&  0.00  & 0.00   & 2.63 & 1.46 & 2.62 &3.03    \\ \hline
     NC 1$\gamma$ &  0.00  & 0.00 & 1.08 & 2.59 & 0.33 &1.49     \\ \hline
     NC Other      &  0.25  & 0.25 & 0.14 & 0.33 & 0.98  &0.18   \\ \hline \hline
     Total  Error     &  4.40  & 3.76 & 6.10 & 6.51 & 20.94  &4.77  \\ \hline \hline
   \end{tabular}
 \end{center}
\end{table}
About the usage of ND280 data to constrain the flux and neutrino interaction model, this analysis incorporates the FGD2  data to include interactions on H$_2$O target. In the previous report in this conference series, only data from FGD1 with carbon target was used. In total, 14 near-detector data samples are used to further constrain the flux and neutrino interaction models, including 6 samples for $\nu$-mode (3 samples per FGD categorized according to number of observed pions (0, 1 and more than 1) in presence of $\mu^{-}$ track) and 8 samples for $\overline{\nu}$-mode (4 samples per FGD divided based on a combination of non-existed pion or existed pion and the presence of $\mu^{+}$ track or $\mu^{-}$ track). To estimate uncertainties on the far-detector events, summarized in Table~\ref{tab:syst}, the near detector systematic and flux parameters are marginalized. %The errors on the number of predicted events in the Super-K samples are . 
%\textcolor{red}{The bottomline is data are used as much as possible at each step to reduce model dependence}
%\subsection{Improvement in analysis}
%The $\sigma_{\nu_{e}}/\sigma_{\nu_{\mu}}$, $\sigma_{\bar{\nu}_{e}}/\sigma_{\bar{\nu}_{\mu}}$ error is a theoretically motivated error based on. For the neutral current (NC) channels, the errors are not constrained by the ND280 data, but based on the theoretical model and external data. 
    % External Constraint on  & \multirow{2}{*}{0.0} & \multirow{2}{*}{0.0} & \multirow{2}{*}{4.2} & \multirow{2}{*}{4.0} & \multirow{2}{*}{0.1} & \multirow{2}{*}{0}    \\ \hhline{~~~~~~~}
  %   $\theta_{12}$, $\theta_{13}$, $\Delta m^{2}_{21}$ & {} & {} & {} & {} & {} & {} \\
   %  \hline \hline
  % \cite{Agashe:2014kda}
A joint fit based on the likelihood maximum approach to five far detector data samples is performed in order to measure the interested oscillation parameters $\sin^2\theta_{23}$, $\Delta m^2_{32}$, $\sin^2\theta_{13}$ and $\delta_{CP}$. T2K experiment is not sensitive to the so-called solar neutrino oscillation parameters, $\Delta m^2_{21}$ and $\theta_{12}$. Priors for for these parameters are taken from PDG, particularly $\sin^22\theta_{12}=0.846\pm 0.021$ and $\Delta m^2_{21} = (7.53\pm 0.18) \times 10^{-5} \text{(eV}^2/\text{c}^4)$. Flat priors are used for $\sin^2\theta_{23}$, $|\Delta m^2_{32}|$ and $\delta_{CP}$. When the reactor measurement is taken into account, a Gaussian prior of $\sin^2\theta_{13} = 0.0857\pm 0.0046$ \cite{Patrignani:2016xqp} is used.
%--------------------figure
\begin{figure}[H]
%\centering
\includegraphics[width=0.32\textwidth]{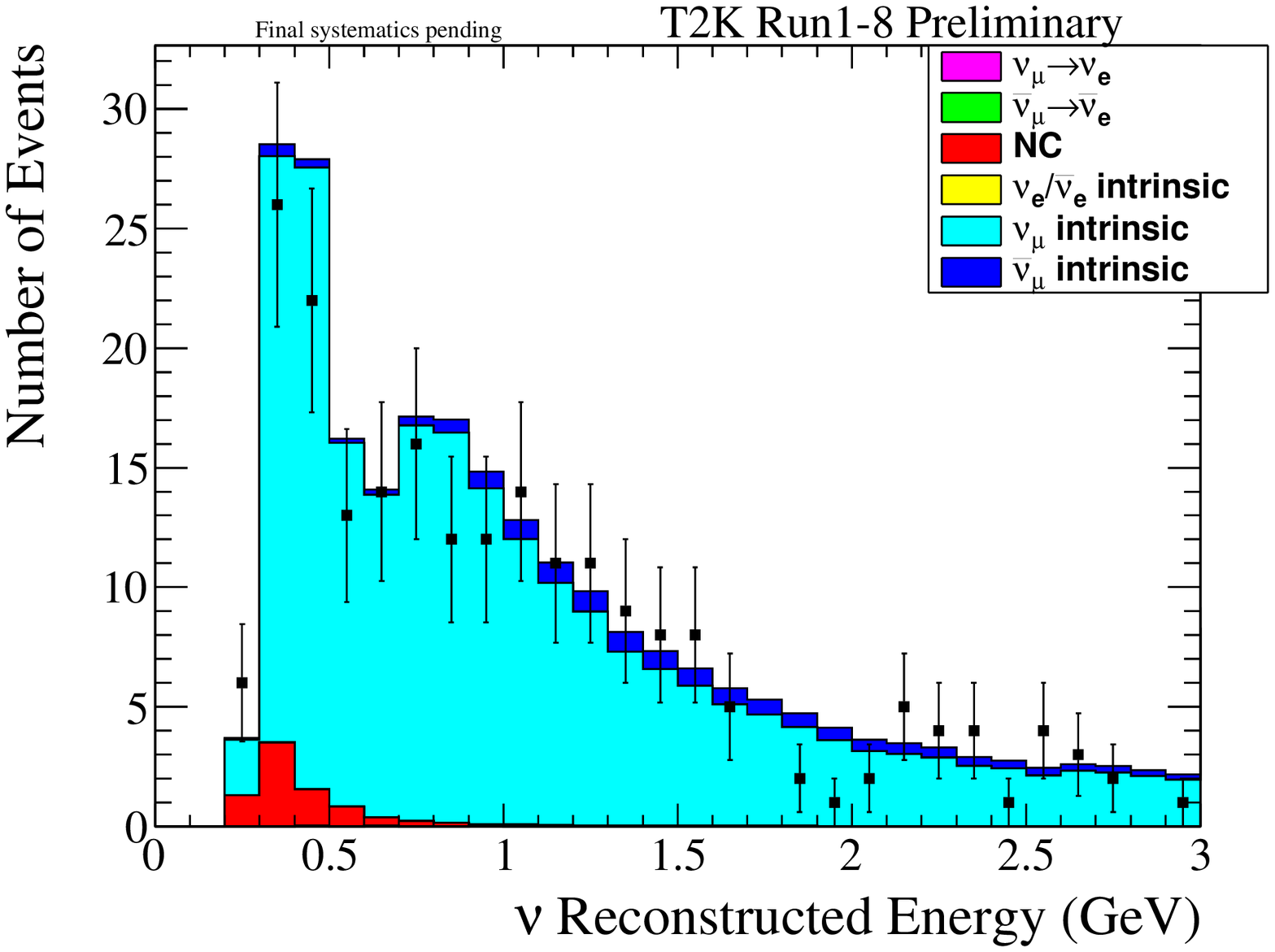} %histo
\includegraphics[width=0.32\textwidth]{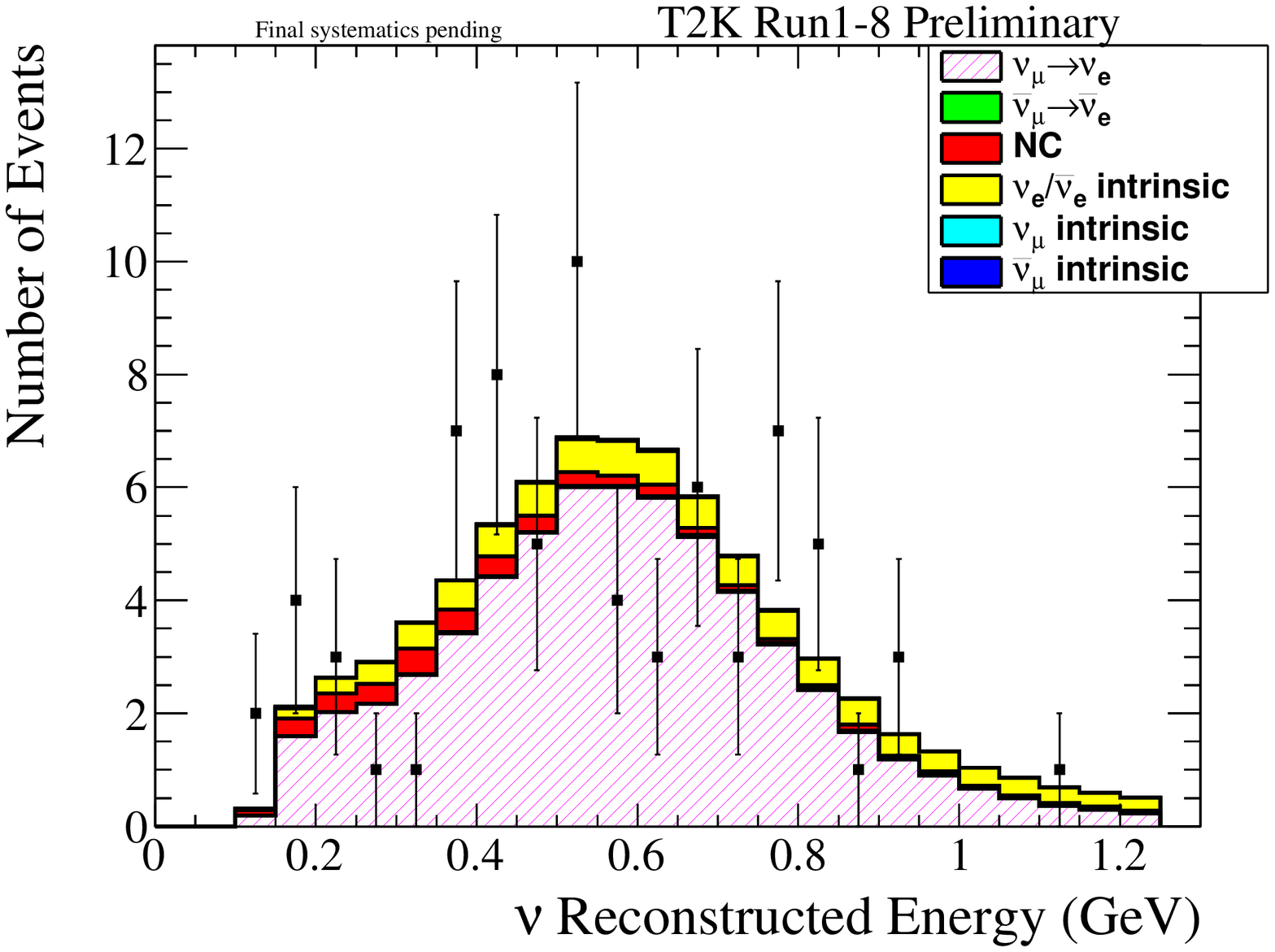} 
\includegraphics[width=0.32\textwidth]{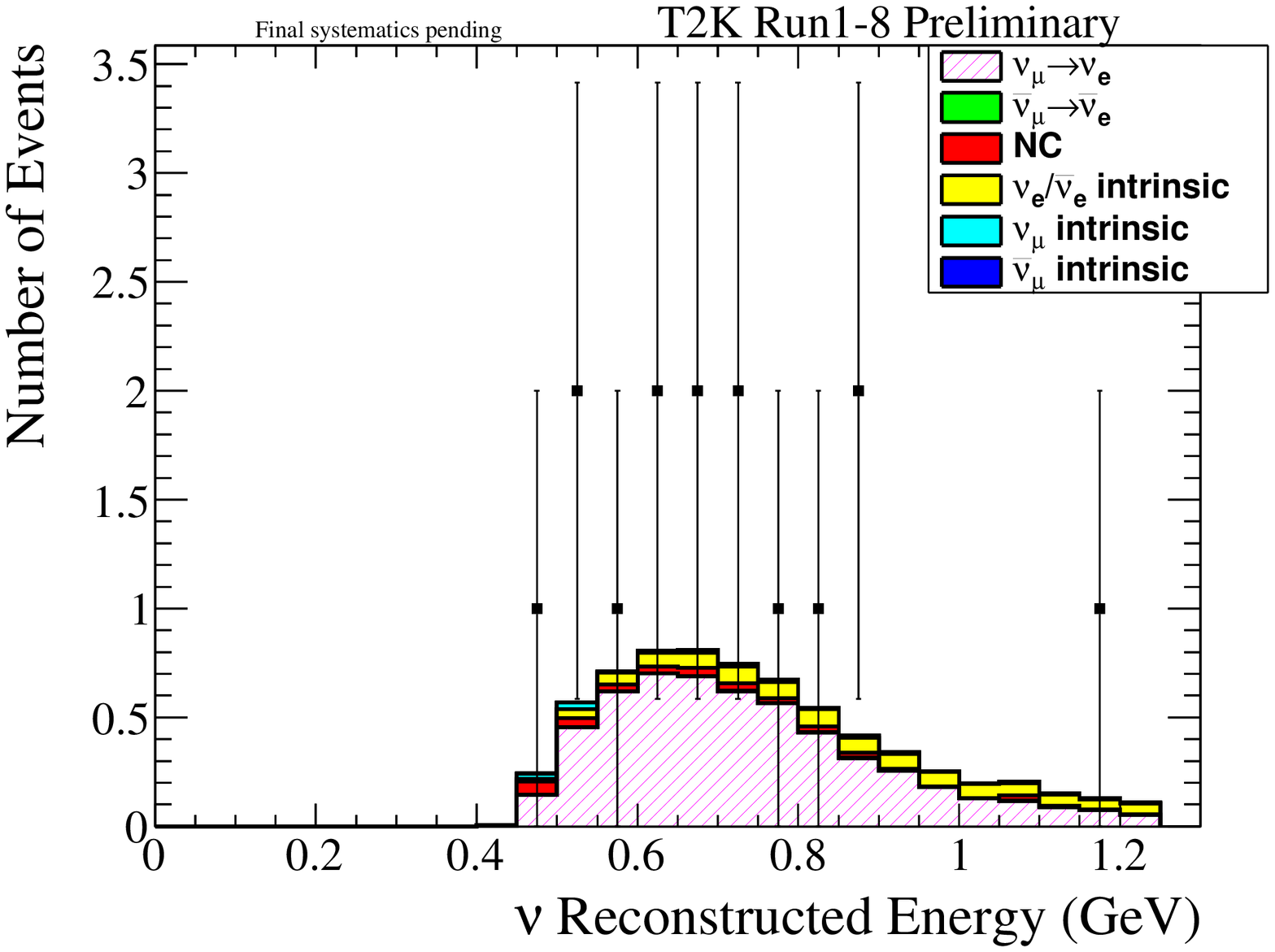} \\
\includegraphics[width=0.32\textwidth]{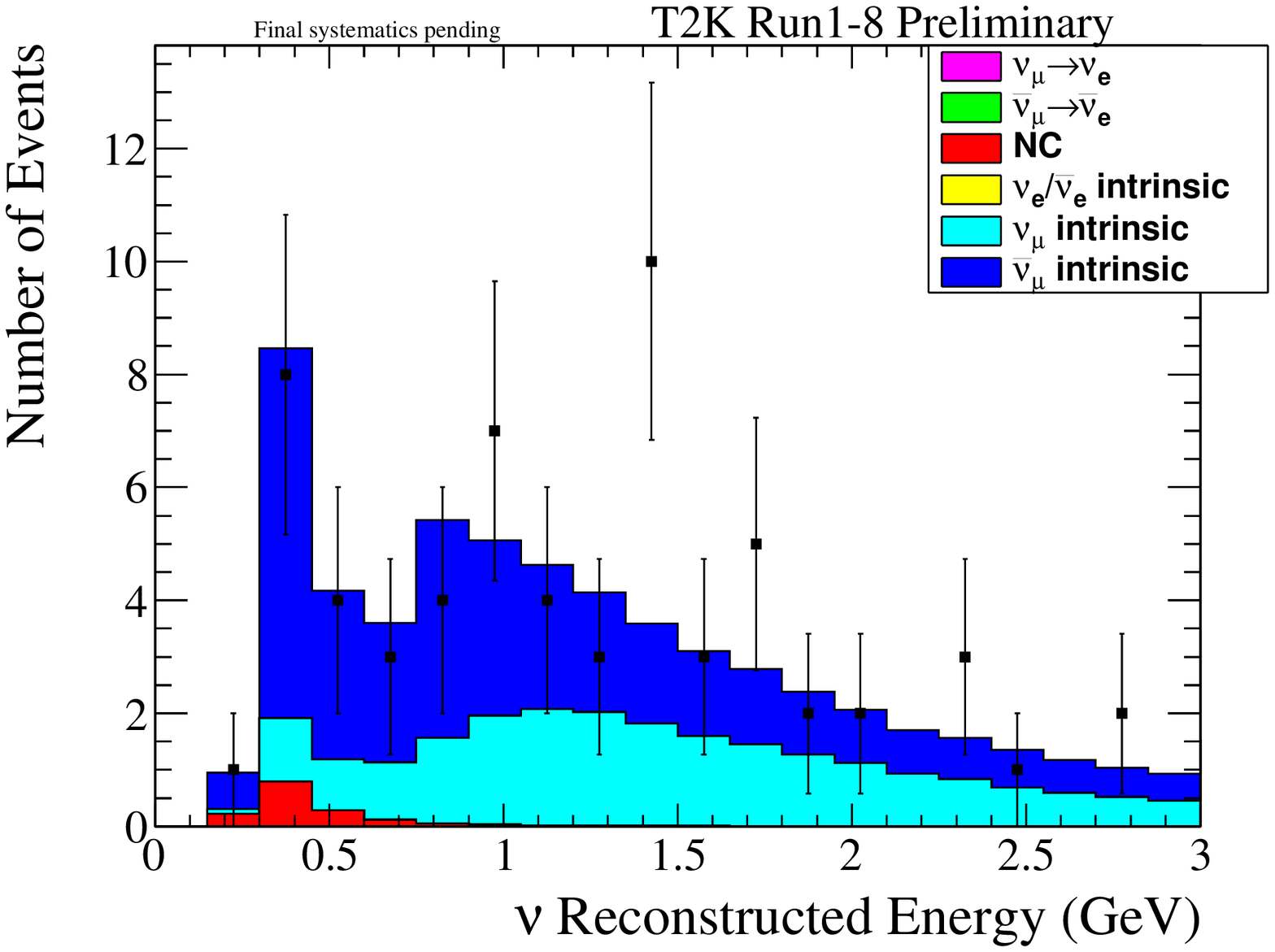} 
\includegraphics[width=0.32\textwidth]{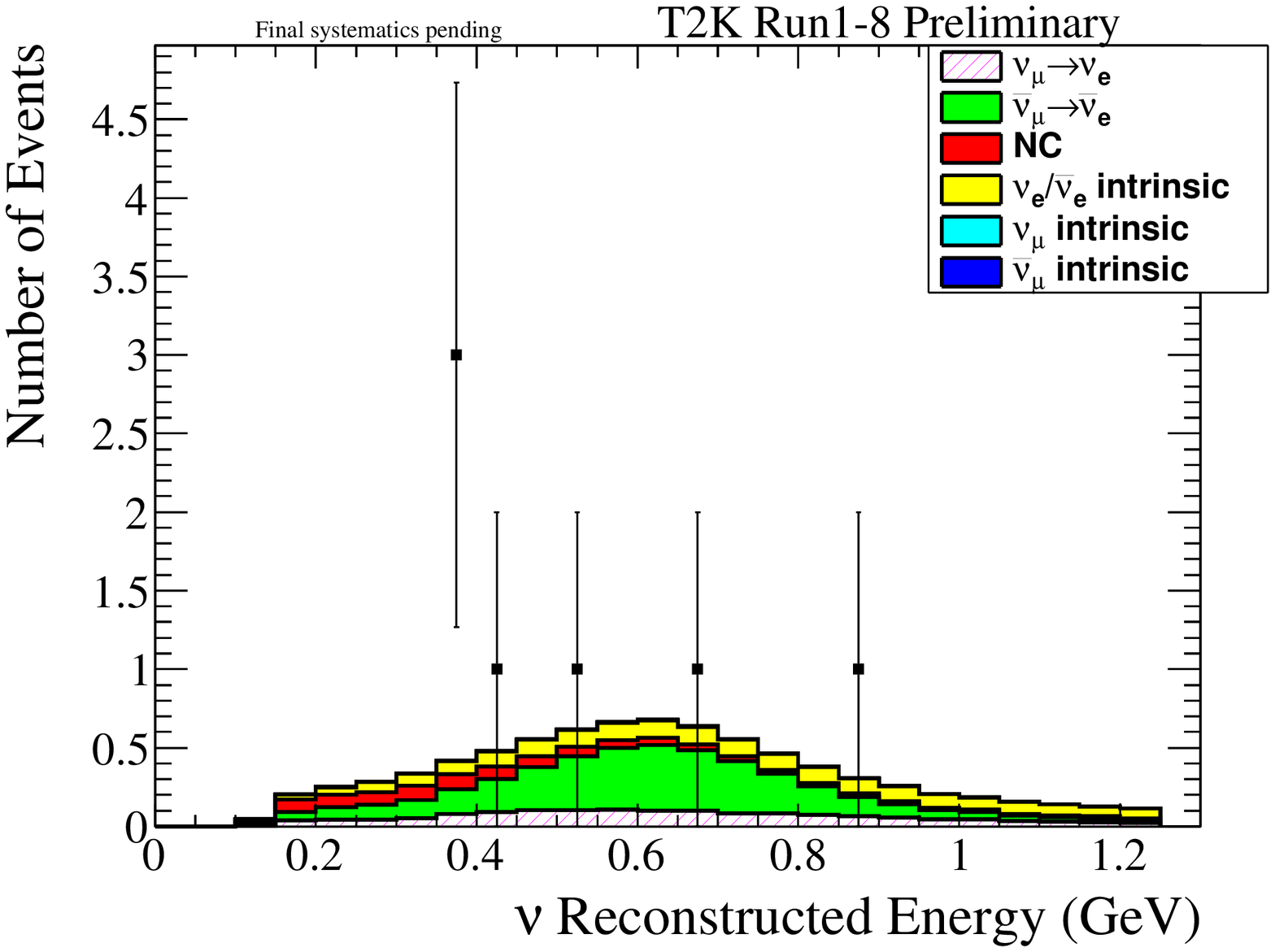}
%histo
\caption{T2K far-detector data sample distributions for $\nu$-mode at top plots (left to right: CCQE-like $\nu_{\mu}$, CCQE-like $\nu_e$ and CC1$\pi^+$-like $\nu_e$) and $\overline{\nu}$-mode at bottom plots (left to right: CCQE-like $\overline{\nu}_{\mu}$ and CCQE-like $\overline{\nu}_{e}$).}
\label{fig:data5sam}
\end{figure}
%--------------------figure

\begin{table}[H]
  \label{tab:evtnum}
    \caption{Observed and predicted number of events in the T2K far-detector data samples. Normal mass hierarchy is assumed.}
\centering
  \begin{tabular}{ | l | c | c | c | c | }
    \hline
    \multirow{2}{*}{Sample} & \multicolumn{3}{|c|}{Prediction at true $\delta_{CP}$} & \multirow{2}{*}{Data}\\ \hhline{~---~}
         & $\delta_{CP}=-\pi/2$ & $\delta_{CP}=0$ & $\delta_{CP}=+\pi/2$ & \\ \hline
    1Ring $e$-like CCQE-enriched, $\nu$-mode  &73.5 & 61.5 & 49.9 & 74\\ \hline
    1Ring $e$-like CC1$\pi^+$-enriched, $\nu$-mode  &6.9 & 6.0 & 4.9 & 15\\ \hline
    1Ring $e$-like CCQE-enriched, $\overline{\nu}$-mode & 7.9 & 9.0 & 10.0 & 7\\ \hline
    1Ring $\mu$-like CCQE-enriched, $\nu$-mode  &267.8 & 247.4 & 267.7 & 240\\ \hline
    1Ring $\mu$-like CCQE-enriched, $\overline{\nu}$-mode  &63.1 & 62.9 & 63.1 & 68\\ \hline
    \hline
  \end{tabular}
\end{table}
%table
%%%%%%%%%%%%%%%%%%%%%%%%%%%%%%%%%%
%%%%%%%%%%%%%%%%%%%%%%%%%%%%%%%%%%
\section{T2K Latest Results and Future Prospects on Neutrino Oscillations}\label{sec:result}
\begin{figure}[H]
\centering
\includegraphics[width=0.45\textwidth]{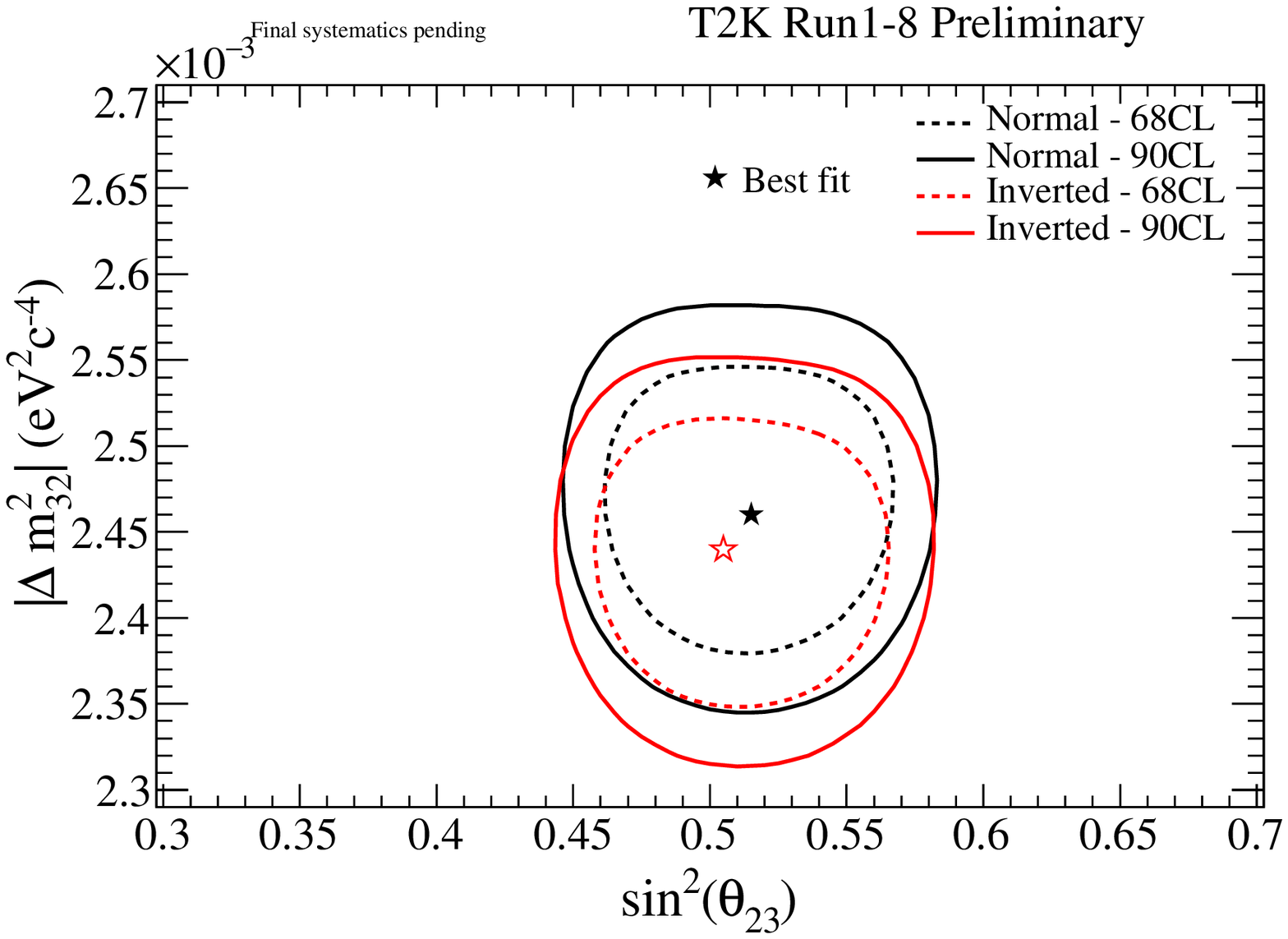} %histo
\includegraphics[width=0.45\textwidth]{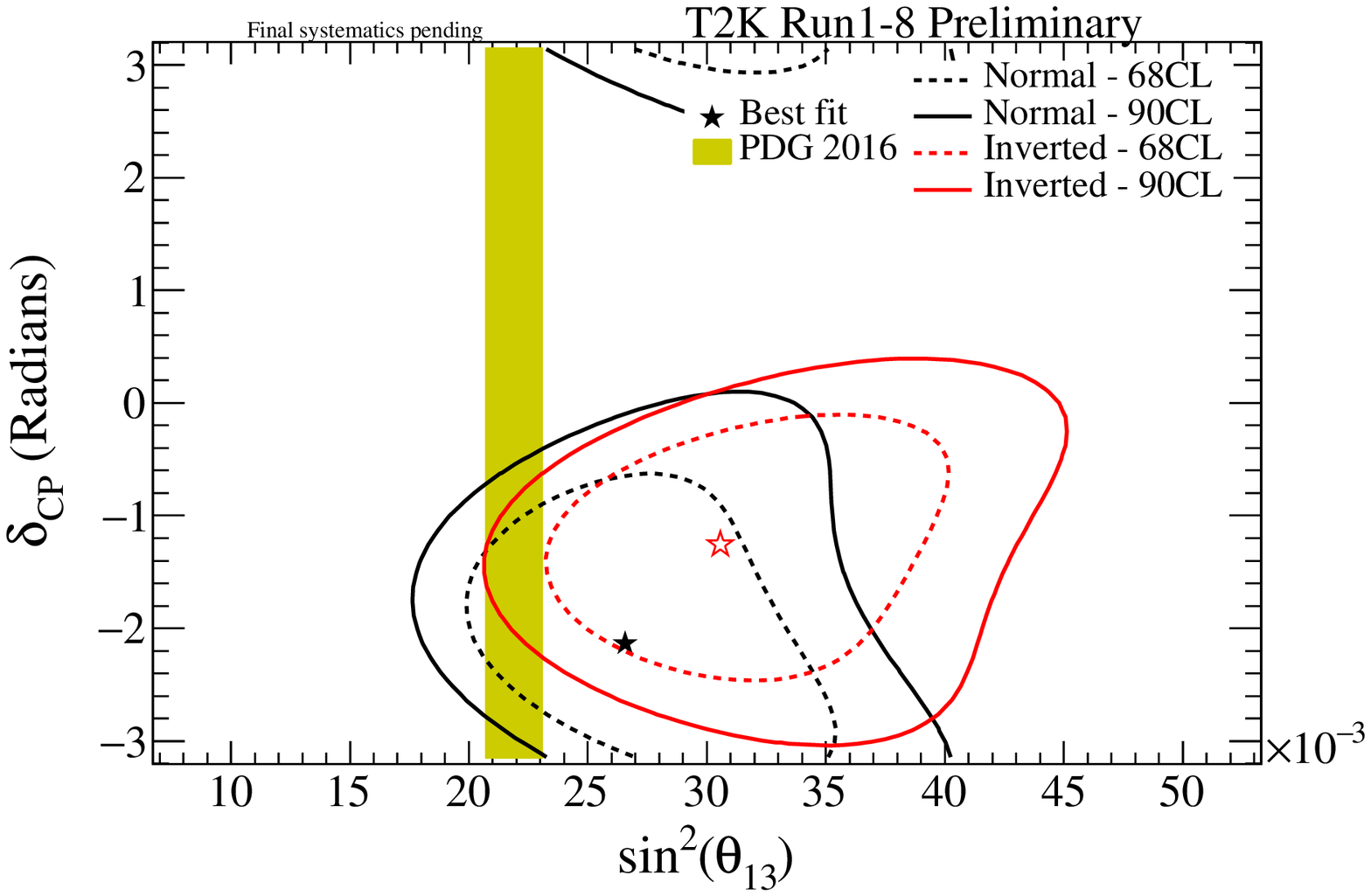} 
%histo
\caption{Left: allowed region for $\Delta m^2_{32}$ vs. $\sin^2\theta_{23}$. Right: favored region on $\sin^2\theta_{13}$ vs. $\delta_{CP}$ overlaid with reactor measurement (yellow band).}
\label{fig:2d}
\end{figure}

\begin{figure}[H]
\begin{minipage}[H]{0.47\linewidth}
\centering
\includegraphics[scale=0.35]{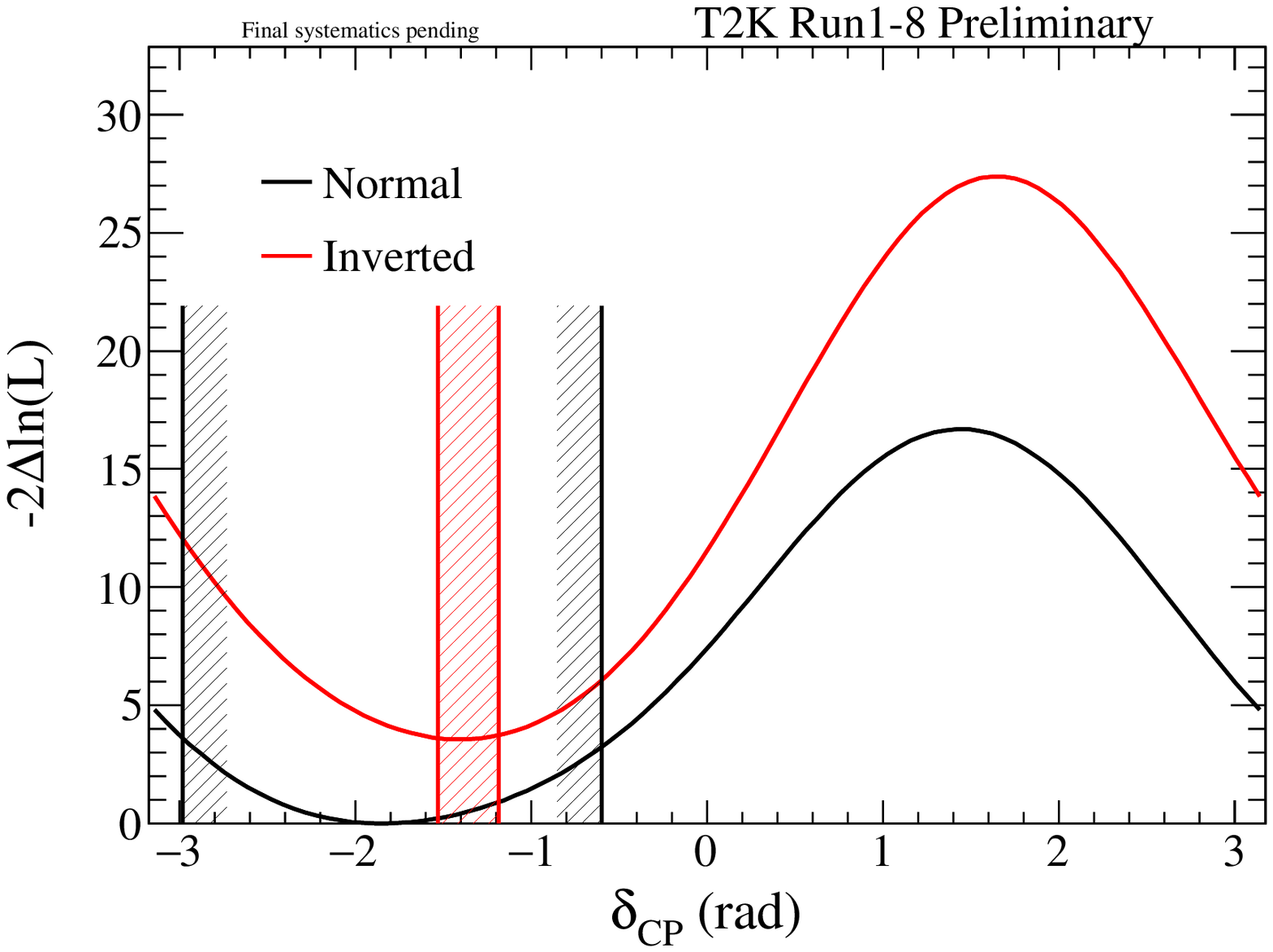} %histo
\caption{ $2\sigma$ confidence interval for the measured $\Delta \chi^2$ distribution. $\theta_{13}$ constrained by reactor measurements is used.}
\label{fig:dcp}
\end{minipage}
\hspace{0.5cm}
\begin{minipage}[H]{0.47\linewidth}
\centering
\includegraphics[scale=0.35]{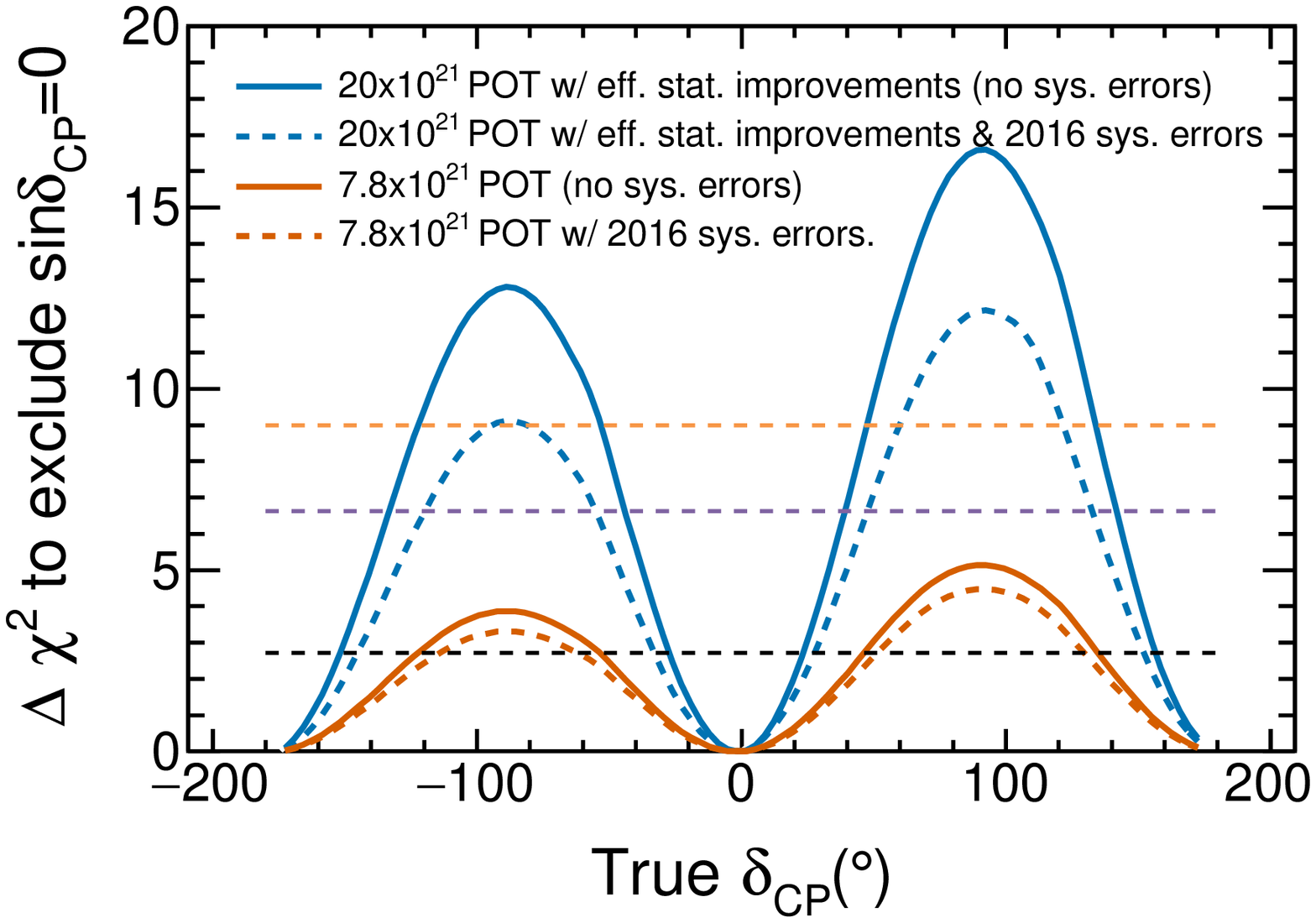} %histo
\caption{Sensitivity of T2K-II as a function of true $\delta_{CP}$ assuming that the mass hierarchy is known. }
\label{fig:future}
\end{minipage}
%\caption{T}
%\label{fig:figure}
\end{figure}

The results presented below are based on a total exposure of $2.23\times 10^{21}$ POT, consisting of $1.47\times10^{21}$ POT in $\nu$-mode and $0.76\times10^{21}$ POT in $\overline{\nu}$-mode. Reconstructed energy distributions of five far-detector data samples with oscillation parameters at best fit values are shown in Fig.~\ref{fig:data5sam}. Number of observed events in T2K far detector in comparison to the MC prediction with various values of $\delta_{CP}$ is shown in Table~\ref{tab:evtnum}. Fig.~\ref{fig:2d} shows allowed region on $\Delta m^2_{32}$ vs. $\sin^2\theta_{23}$ and favored region on $\sin^2\theta_{13}$ vs. $\delta_{CP}$. T2K data continue to favor the maximal disappearance. By marginalizing the likelihood over the nuisance parameters and other oscillation parameters, the resulting posterior probabilities show a moderate preference to normal mass hierarchy (86.8$\%$) and $\theta_{23}>45^{\circ}$ (78.1$\%$). Parameter $ \sin^2\theta_{13} = 0.0277^{+0.0054}_{-0.0047}$, measured by T2K data, is consistent with reactor measurements. Fig.~\ref{fig:dcp} shows 2$\sigma$ C.L. interval for the measured $\Delta \chi^2$ distribution as a function of $\delta_{CP}$ which is obtained by integrating over other parameters and taking the reactor measurement into account. The best fitted values of $\delta_{CP}$ are closed to $-\pi/2$ for both cases of mass hierarchy. The resulting intervals at 2$\sigma$ C.L. are [-2.91,-0.60] for normal mass hierarchy ($\Delta m^2_{32}>0$) and [-1.54, -1.19] for inverted mass hierarchy ($\Delta m^2_{32}<0$). The CP conserving values (0 and $\pi$) fall outside of these intervals.
%%%https://www.t2k.org/docs/plots/045/oa-plots/valorofficialplots
%--------------------figure
%--------------------figure
%---------------figure
%------------------figure
%\begin{align*}
%\text{ (T2K):}  \sin^2\theta_{13} = 0.0277^{+0.0054}_{-0.0047} \text{ (PDG 2016):  } \sin^2\theta_{13} = 0.0210 \pm 0.0011 
%\end{align*}
%--------------------figure
%\begin{figure}[H]
%\centering
%\includegraphics[width=0.47\textwidth]{deltacp_interval_2sigma.eps} %histo
%\includegraphics[width=0.47\textwidth]{elike_combined_dcp_0_coarse.pdf} 
%histo
%\caption{ Left: $2\sigma$ confidence interval for the measured $\Delta \chi^2$ distribution. Right: distributions of observed and predicted at $\delta_{CP}=0$ events in comparing appearance channels between $\nu$-mode and $\overline{\nu}$-mode.}
%\label{fig:dcp}
%\end{figure}
%--------------------figure
%For 1$\sigma$ C.L.  interval [-2.49, -1.23] (rad.). For 2$\sigma$ C.L. confidence interval
%\begin{align*}
%\text{  Normal MH: } [-2.91,-0.60]; \text{  Inverted MH: } [-1.54, -1.19]
%\end{align*}
%
%\section{T2K Future Prospects}

Approved T2K statistics, $7.8\times 10^{21}$ POT, can be accumulated by 2021. J-PARC aims for upgrade and operations at and higher than 1MW from 2021. Hyper-Kamiokande, the next generation of neutrino experiments in Japan, is expected to start operation around 2026. T2K-II, extended operation until 2026, will collect 20$\times 10^{21}$ POT. Such amount of data along with neutrino beamline upgrade and analysis improvement make T2K(-II) physics potentials even more interesting. In particular, 3$\sigma$ or higher significance sensitivity to CP violation can be achieved if $\delta_{CP}$ close to $\pi/2$, as shown in Fig.~\ref{fig:future}. The precision of $\Delta m^2_{32}$ can be 1$\%$ while that on $\theta_{23}$ is about $0.5^{\circ}$-1.7$^\circ$ depending on the true value. For more detail, refer to  physics potential of T2K-II \cite{Abe:2016tez}. Also it is shown in Fig.~\ref{fig:future} that the impact on $\delta_{CP}$ measurement from systematic error is significant and thus motivates for the ND280 upgrade which is now happening actively.\\

\end{document}